\documentclass[a4paper,12pt]{article}
\usepackage[utf8]{inputenc}
\usepackage{amsmath}
\usepackage{amssymb}
\usepackage{bbold}
\usepackage{amsfonts,bm}
\usepackage{geometry}
\usepackage{fullpage}
\usepackage{graphicx}
\usepackage{caption}
\usepackage{todonotes}
\usepackage{multicol}
\usepackage{multirow}
\usepackage{subcaption}
\usepackage{slashed}
\usepackage{float}
\usepackage{color,xcolor}
\usepackage{epstopdf}
\usepackage{units}
\usepackage[normalem]{ulem} 
\usepackage{jheppub} 
\usepackage{bm,paralist,xspace,url,relsize}
\usepackage[]{hyperref}

\newcommand{\fac}{\ensuremath{\text{FACET}}\xspace}
\newcommand{\fas}{\ensuremath{\text{FASER2}}\xspace}

\title{Sensitivity of the FACET experiment to Heavy Neutral Leptons and Dark Scalars}
 \author[a, b]{Maksym~Ovchynnikov,}
 \author[c]{Viktor~Kryshtal,}
 \author[d,e,f]{Kyrylo~Bondarenko}
  \affiliation[a]{Instituut-Lorentz, Leiden University, Niels Bohrweg 2, 2333 CA Leiden, The Netherlands}
  \affiliation[b]{Institut für Astroteilchen Physik, Karlsruher Institut für Technologie (KIT), Hermann-von-Helmholtz-Platz 1, 76344 Eggenstein-Leopoldshafen, Germany
}
  \affiliation[c]{Department of Physics, Taras Shevchenko National University of Kyiv, 64 Volodymyrs’ka str., Kyiv 01601, Ukraine}
  \affiliation[d]{
IFPU, Institute for Fundamental Physics of the Universe, via Beirut 2, I-34014 Trieste, Italy}
\affiliation[e]{
SISSA, via Bonomea 265, I-34132 Trieste, Italy
}
\affiliation[f]{
INFN, Sezione di Trieste, SISSA, Via Bonomea 265, 34136, Trieste, Italy
}

\emailAdd{maksym.ovchynnikov@kit.edu}
\emailAdd{victor.kryshtal@gmail.com}
\emailAdd{kyrylo.bondarenko@sissa.it}

\date{}

\begin{document}
\abstract{We analyze  the potential of the recently proposed experiment FACET (Forward-Aperture CMS ExTension) to search for new physics. As an example, we consider the models of Higgs-like scalars with cubic and quartic interactions and Heavy Neutral Leptons. We compare the sensitivity of FACET with that
of other proposed ``intensity frontier'' experiments, including  FASER2, SHiP, etc. and demonstrate that FACET could probe an interesting parameter space between the current constraints and the potential reach of the above mentioned proposals.
}

\maketitle

\section{Introduction}
Despite its success in describing accelerator data, the Standard model (SM) fails to explain several observed phenomena constituting beyond the Standard model problems: neutrino masses, dark matter, and the matter-antimatter asymmetry. These problems may be resolved by extending the SM particle content with some new particles. One class of extensions is where new particles interact with SM via renormalizable operators suppressed by very small couplings, the so-called portals. Depending on the spin of the mediator field entering the portal operator, there are three types of portals -- scalar, vector, and fermion~\cite{Alekhin:2015byh,Agrawal:2021dbo}.

To search for portal particles, many experiments have been proposed during the last few years. Examples include dedicated beam experiments such as SHiP~\cite{SHiP:2015vad}, DUNE~\cite{DUNE:2015lol}, SHADOWS~\cite{Baldini:2021hfw}, NA62~\cite{NA62:2017rwk}; LHC-based experiments, such as MATHUSLA~\cite{Curtin:2018mvb}, Codex-b~\cite{Aielli:2019ivi}, ANUBIS~\cite{Bauer:2019vqk}, AL3X~\cite{Gligorov:2018vkc}. There is a class of LHC-based experiments that have decay volume covering large pseudorapidities, which is called the far-forward experiments. The particles produced in the far-forward direction have large energies, $E= \mathcal{O}(1\text{ TeV})$, which means that their lifetime is increased by $\gamma \sim 10^{3}(1\text{ GeV}/m)$. Therefore, as compared with the off-axis experiments located at the same distance, the far-forward experiments may probe particles with shorter lifetimes. 

The representatives of this class are already running FASER~\cite{FASER:2018bac,FASER:2019aik}, FASER$\nu$~\cite{FASER:2019dxq,FASER:2020gpr} and SND@LHC~\cite{SHiP:2020sos} experiments. Their proposed upgrades, FASER2/FASER$\nu$2 and AdvSND, would be installed at the Far Forward physics facility and work during the High Luminosity phase of the LHC~\cite{Feng:2022inv}. Recently, a new far-forward experiment FACET has been proposed~\cite{Cerci:2021nlb}. Apart from covering $\simeq 4$ times larger solid volume and having longer decay volume, it would be located in 100 meters downwards the CMS interaction point -- $\simeq 5$ times closer than SND@LHC/FASER, and in this way allows to probe even shorter lifetimes~\cite{Du:2021cmt,Liu:2022ugx}. 

\begin{figure}[!h]
    \centering
    \includegraphics[width=0.5\textwidth]{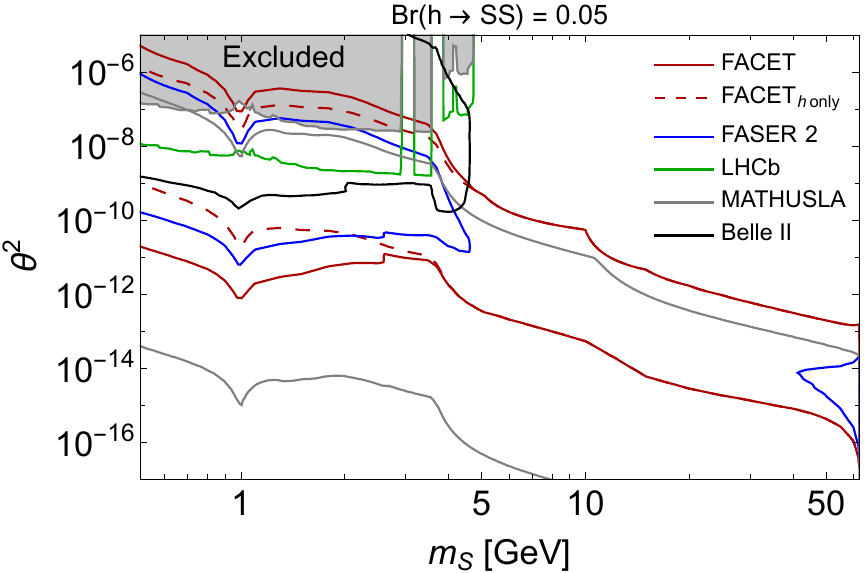}~\includegraphics[width=0.5\textwidth]{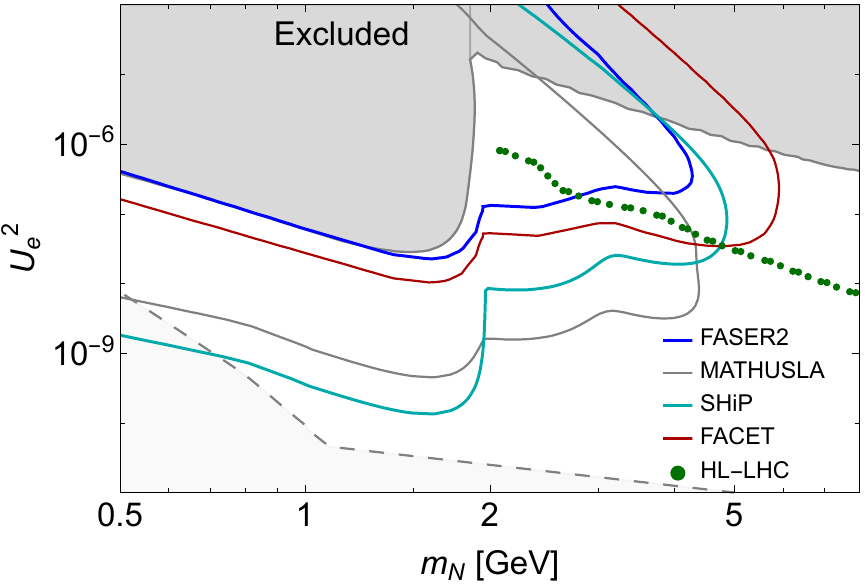}
    \caption{Sensitivity of \fac and \fas to the models of HNLs and Higgs-like scalars. Left panel: Higgs-like scalars, assuming $\text{Br}(h\to SS)=0.05$. The solid red line shows the sensitivity of \fac including the production of scalars from $h$ and $B$, while the dashed line denotes the sensitivity to scalars from $h$ only. We also include the sensitivity of Belle II from~\cite{Kachanovich:2020yhi} (see also~\cite{Filimonova:2019tuy}). Right panel: HNLs that mix predominantly with $\nu_{e}$. For comparison, we also show the sensitivity of SHiP and MATHUSLA experiments from~\cite{Agrawal:2021dbo}, as well as the optimistic estimate of the sensitivity of HL-LHC from~\cite{Drewes:2019fou}. The region excluded by the previous experiments is given from~\cite{Abdullahi:2022jlv} for HNLs and from~\cite{Agrawal:2021dbo} for Higgs-like scalars.}
    \label{fig:sensitivity}
\end{figure}

In this work, we estimate the sensitivity of the \fac experiment to models of scalar and fermion portals, making a qualitative comparison of its sensitivity with \fas. The final results are shown in Fig.~\ref{fig:sensitivity}, where we also show the sensitivities of other proposed experiments such as SHiP, MATHUSLA, Belle II, and LHC, to demonstrate the possible synergy between these searches. We see that due to larger decay volume and closer distance from the interaction point, \fac allows to significantly extend the probed parameter space as compared to \fas.

The content of the paper is as follows. In Sec.~\ref{sec:portals}, we briefly describe the scalar and neutrino portals. In Sec.~\ref{sec:facet}, we describe the FACET experiment. In Sec.~\ref{sec:analytic-estimates}, we compare the reach of the FACET and FASER2 experiments based on semi-analytic estimates, considering scalars and heavy neutral leptons produced by decays of $B$ mesons and Higgs bosons. In Sec.~\ref{sec:sensitivity-summary}, we discuss the obtained results and make their comparison with the literature. Finally, in Sec.~\ref{sec:conclusions}, we make conclusions.

\section{Portals}
\label{sec:portals}
\subsection{Scalar portal with the quartic coupling}
The general form of the Lagrangian of the scalar portal below the electroweak scale~\cite{Alekhin:2015byh} is
\begin{equation}
    \mathcal{L}_{\text{scalar portal}} = \theta m_{h}^{2}h S + \frac{\alpha}{2} h SS,
    \label{eq:scalar-portal}
\end{equation}
where $h$ is the Higgs boson field, $S$ is a new scalar particle (also called the Higgs-like scalar, or dark scalar), $\theta$ is the mixing angle, and $\alpha$ is the quartic coupling constant. The scalar with mass in GeV range may be a mediator between the SM and dark matter particles or serve as a light inflaton~\cite{Bezrukov:2009yw}. The phenomenology of the scalar portal at accelerator experiments has been intensively studied in~\cite{Boiarska:2019jym,Bird:2004ts,Batell:2009jf,Bezrukov:2009yw,Clarke:2013aya,Schmidt-Hoberg:2013hba,Evans:2017lvd,Bezrukov:2018yvd,Monin:2018lee,Winkler:2018qyg,Frugiuele:2018coc,Helmboldt:2016zns} as well as in~\cite{Voloshin:1985tc,Raby:1988qf,Truong:1989my,Donoghue:1990xh,Willey:1982ti,Willey:1986mj,Grzadkowski:1983yp,Leutwyler:1989xj,Haber:1987ua,Chivukula:1988gp} in the context of the light Higgs boson.

At the LHC, Higgs bosons are copiously produced. In particular, during the high luminosity phase, around $10^{8}$ bosons will be generated. In the case $\alpha \neq 0$, they may decay into a pair of scalars through the process $h\to SS$. 

Current constraints on $\alpha$ are not very restrictive for the model of scalars. Indeed, the strongest bound on $\alpha$ comes from searches for invisible decays $h\to \text{inv}$ at ATLAS and CMS, constraining $\text{Br}(h\to \text{inv}) < 0.15$~\cite{Sirunyan:2018owy,Aaboud:2018sfi}. During the high luminosity phase of the LHC, it would be possible to probe the branching ratio down to the values $\text{Br}(h\to \text{inv}) = 0.05$~\cite{Bechtle:2014ewa}. 

The number of Higgs bosons that would be produced during the High Luminosity phase of the LHC is $N_{h}\simeq 2\cdot 10^{8}$. Therefore, given the current constraints on $\text{Br}(h\to \text{inv})$, the production channel $h\to SS$ allows to significantly extend the reach of the LHC and LHC-based experiments, making it possible to search for the scalars with masses up to $m_{S} \simeq m_{h}/2$.

The parameter space of dark scalars excluded by past experiments and probed by proposed LHC experiments (we choose \fas and MATHUSLA as a representative example), assuming $\text{Br}(h\to SS) = 0.05$, is shown in the left panel of Fig.~\ref{fig:sensitivity}. In the paper~\cite{Boiarska:2019vid}, it has been demonstrated that \fas has a limited potential to probe this model, mainly due to the suppressed small angular coverage and short length of the decay volume.

\subsection{Heavy Neutral Leptons}
The Lagrangian of the fermion portal is
\begin{equation}
    \mathcal{L}_{f} = F_{\alpha I}\bar{L}_{\alpha}\tilde{H}\mathcal{N}_{I}+\text{h.c.} + \ \mathcal{N}\text{ mass term},
    \label{eq:mass-mixing}
\end{equation}
where $\mathcal{N}_{I}, i = 1,2,\dots$ is a massive fermion (that may be either Dirac or Majorana depending on the $\mathcal{N}$ mass term), $\tilde{H} = i\sigma_{2}H^{*}$ is the Higgs doublet in the conjugated representation, $L_{\alpha}$ is the SM lepton doublet ($\alpha = e,\mu,\tau$), and $F_{\alpha I}$ are complex couplings. Below the scale of the electroweak symmetry breaking, the first term in~\eqref{eq:mass-mixing} induces a mass mixing between the $\mathcal{N}$ and active neutrinos. The mixing is parametrized by the mixing angle $U_{\alpha I} = F_{\alpha I}/\sqrt{2}v \ll 1$, where $v$ is the Higgs VEV. As a result, the combination of active neutrinos $\sum_{\alpha}F_{\alpha I}$ and the fermion $\mathcal{N}_{I}$ are a combination of two mass eigenstates -- a very light neutrino and a heavy neutral lepton $N_{I}$ (HNL). 

The mass mixing determines the way how HNLs interact with SM particles. Similarly to the interaction of the SM active neutrinos, it is with other neutrinos and charged leptons via $W$ and $Z$ bosons. The only difference is that the HNL couplings are suppressed by $U_{\alpha I}$.

The parameter space of HNLs is shown in the right panel of Fig.~\ref{fig:sensitivity}. We show SHiP, FASER2, MATHUSLA, and high luminosity LHC among the proposed experiments. At the LHC, HNLs heavier than kaons may be produced in decays of D, B mesons, and W bosons. The last channel allows extending the maximal HNL mass reach from $m_{N} = m_{B}\simeq 5\text{ GeV}$ to $m_{N}\simeq m_{W}$ as compared to the dedicated beam experiments such as SHiP. However, this is not the case for either MATHUSLA or \fas, since they are located too far from the HNL production point, and the HNLs produced from $W$ bosons in the accessible parameter space are too short-lived to reach the decay volume~\cite{Bondarenko:2019yob}.

\section{FACET experiment}
\label{sec:facet}
FACET (Forward Aperture CMS ExTension)~\cite{{Cerci:2021nlb}} is a recent proposal of a subsystem of CMS to be added to search for long-lived particles during the High Luminosity (HL) phase of the LHC.

\begin{figure}
    \centering
    \includegraphics[width=0.7\textwidth]{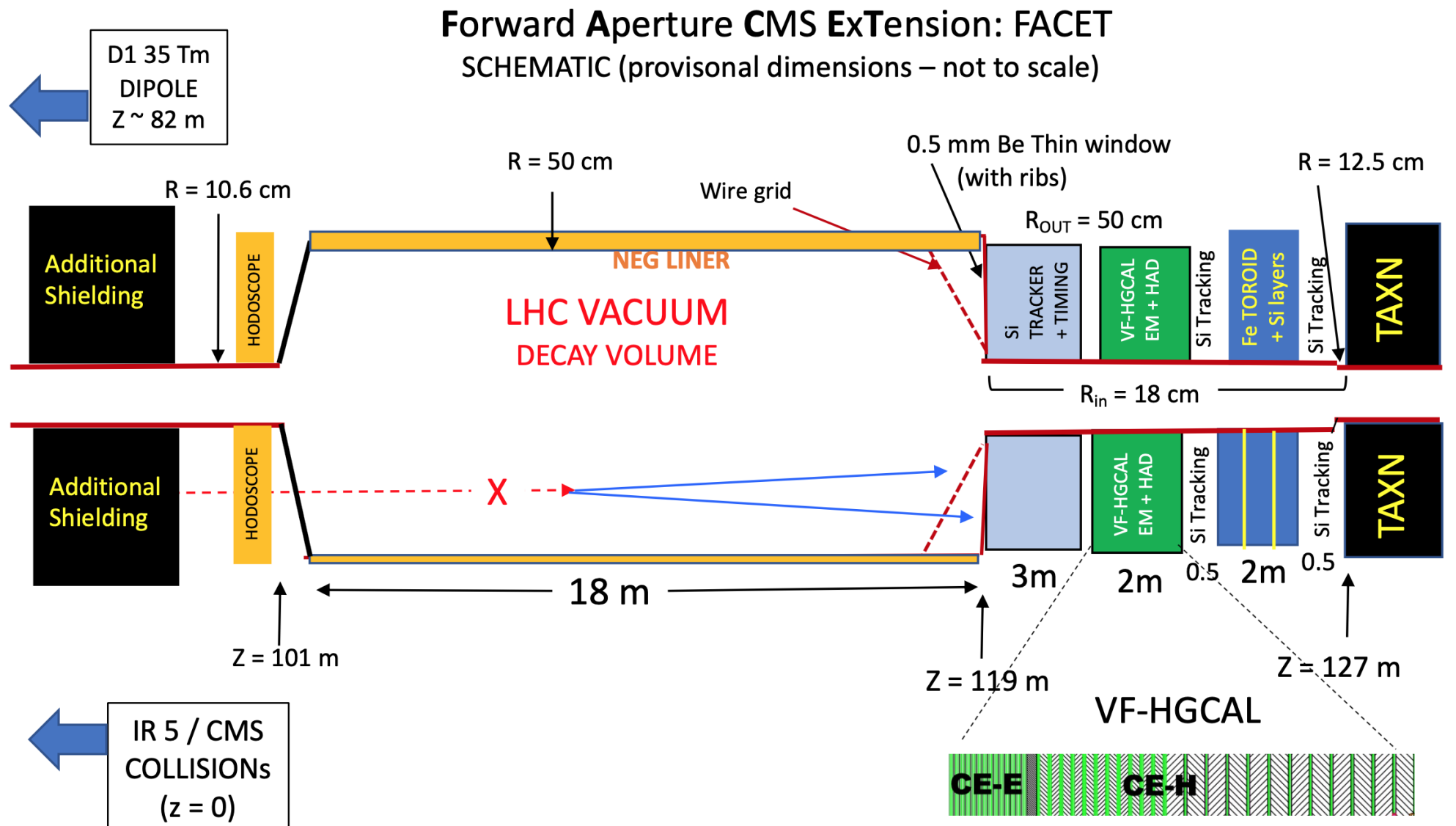}
    \caption{The schematic layout of the FACET experiment (see text for details). The figure is taken from~\cite{Cerci:2021nlb}.}
    \label{fig:FACET}
\end{figure}

The schematic layout of FACET is shown in Fig.~\ref{fig:FACET}. Given $z$ as the distance from the CMS experiment along the beam axis, FACET will be located between the 35 T$\cdot$m superconducting beam separation dipole D1 at $z = 80$~m and the TAXN absorber at $z = 128$~m. The decay volume is an enlarged proton beam pipe with radius $r = 0.5$~m located from $z = 101\text{ m}$ to $z = 119\text{ m}$. 

The detector part is located right after the decay volume. It has shape of the annulus with the inner radius $r_{\text{in}} = 18\text{ cm}$ and outer radius $r_{\text{out}} = 50\text{ cm}$. The detector covers polar angles $1.5<\theta <4\text{ mrad}$ and consists of $\simeq 3$ m of silicon tracker with the transversal resolution of $\sigma_{x,y} = 30\ \mu\text{m}$, the timing layer with Low-Gain Avalanche Detectors (LGAD) having resolution $\sigma_{t} \sim 30\text{ ps}$, and a high granularity EM and hadronic calorimeter. 

The background is greatly reduced because of 200-300 $\lambda_{\text{int}}$ of magnetized iron in the LHC quadrupole magnets Q1–Q3. Detailed FLUKA simulations predict $\simeq 30$ charged particles with momentum $p>1\text{ GeV}$ and $\simeq 1$ light neutral hadron ($K^0_S, K^0_L, \Lambda$) per bunch crossing~\cite{Cerci:2021nlb}. Charged particles and decays of neutral hadrons may mimic decays of new physics particles. The combination of the precision hodoscope (with the inefficiency of $10^{-5}$) and precision tracking reduces the background for most of the new physics decay channels down to a negligible level for the mass of a new physics particle $m_X \gtrsim 0.8\text{ GeV}$. Nevertheless, decays of neutral hadrons create a background in the region $m_X \lesssim 0.8\text{ GeV}$, making searching for new particles in this mass range complicated.

However, given the specific model, the background may be greatly reduced. First, one may utilize specific final states of decays of the HNLs and dark scalars, see~\cite{Bondarenko:2018ptm,Boiarska:2019jym} for detail. In particular, the neutral SM particles do not decay (or decay very rarely) into solely a dilepton pair~\cite{Workman:2022ynf}. In contrast, these may be the main decay modes of light scalars (the decay $S\to l^{+}l^{-}$) and the HNLs (decays $N\to l^{+}l^{'-}\nu$, $N\to \pi^{+}l^{-}$). The only caveat is the situation when the decay product would be falsely reconstructed (in particular). In addition, in the case of dark scalars and HNLs with masses up to $\mathcal{O}(1\text{ GeV})$, most of their decays (into $l^{+}l^{-}, \pi^{+}\pi^{-}$, $N\to \pi^{+}l^{-}$) -- are fully reconstructable. Therefore, the invariant mass distribution is peaked at the true scalar/HNL mass. Collecting even a few of such events would be enough to distinguish them from similar events from decays of neutral particles, for which the distribution is peaked at $m_{K^{0}}$ (a very rare process $K\to \pi^{+}\pi^{-}$) or is continuous (the decay $K\to \pi^{-}l^{+}\bar{\nu}$).

\section{\fac vs \fas: qualitative comparison of the sensitivity}
\label{sec:analytic-estimates}
In this section, we compare the sensitivities of the \fac and \fas experiments. For this purpose, we will use semi-analytic estimates similar to the ones presented in~\cite{Bondarenko:2019yob}.

We estimate the number of events with decays of a particle $Y = S,N$ at the \fac and \fas experiments (see Table~\ref{tab:parameters}) in the following way:
\begin{multline}
    N_{\text{events}} = \sum_{X}N_{X}\times \chi_{S/N}^{(X)}\cdot\text{Br}(X\to Y)\times \\ \times \int \frac{dP_{\text{decay}}(\gamma_{Y},z)}{dz} f_{\theta_{Y},E_{Y}} \cdot \epsilon_{\text{decay}}(\gamma_{Y},\theta_{Y},z) d\theta_{Y}dE_{Y}dz
    \label{eq:nevents}
\end{multline}
Here, $X$ corresponds to the decaying particle that produces scalars, with $N_{X}$ being the total number of SM particles produced during the high luminosity phase of the LHC. We assume $N_{h} = 1.7\cdot 10^{8}$ from~\cite{Cepeda:2019klc}, take $N_{B} = 2.4\cdot 10^{15}$, $N_{D} = 5\cdot 10^{16}$, from FONLL~\cite{Cacciari:1998it,Cacciari:2001td,Cacciari:2012ny,Cacciari:2015fta} at the upper bound of uncertainties (see a discussion in~\cite{Bondarenko:2019yob}), and $N_{W} = 3.3\cdot 10^{11}$ from~\cite{ATLAS:2016nlr}. 

$\chi_{Y}^{(X)} = 1$ or $2$ is the number of particles $Y$ produced per decay of $X$.

The integration in~\eqref{eq:nevents} is performed over $Y$ energies $E_{Y}$, polar angles $\theta_{Y}$, and the distance from the collision point along the beam axis $z$ within $l_{\text{min}}<z<l_{\text{min}}+l_{\text{fid}}$, where $l_{\text{min}}$ is the distance to the beginning of the decay volume, and $l_{\text{fid}}$ is the decay volume length. $dP_{\text{decay}}/dz = \frac{e^{-z/c\tau_{Y}\gamma_{Y}}}{c\tau_{Y}\gamma_{Y}}$ is the differential decay probability. 

$f^{(X)}_{\theta_{Y},E_{Y}}$ is the angle-energy distribution of $Y$ produced by decays of $X$. To derive it, we have followed the semi-analytic approach summarized in~\cite{Boiarska:2019vid}; namely, we have integrated the differential distribution $d\text{Br}(X\to Y)$ multiplied with the distribution of the mother particle $X$ over $X$ energy and angles. The mother particle distributions have been obtained from FONLL~\cite{Cacciari:1998it,Cacciari:2001td,Cacciari:2012ny,Cacciari:2015fta} ($B,D$), by the method described in~\cite{Boiarska:2019vid} (for $h$), and from~\cite{Kling:2021fwx} (for $W$).

$\epsilon_{\text{decay}}(\gamma_{S},\theta_{S},z)$ is the decay acceptance -- the fraction of decay products from $Y$ intersecting the front plane of the detector. We estimate it using a simple Monte Carlo simulation by approximating the decay of scalars into two massless particles, and of HNLs into three massless particles via the charged current.\footnote{The approximation works within 25\% accuracy for the whole scalar/HNL mass range.}

Further, we will assume that both \fac and \fas are background-free experiments. Taking into account considerable background from neutral hadron decays on \fac for $m_Y \lesssim 0.8\text{ GeV}$, discussed in Section~\ref{sec:facet}, the obtained results for \fac are only valid above this mass. Parameters of the experiments are summarized in Table~\ref{tab:parameters}.

\begin{table}[h!]
    \centering
    \begin{tabular}{|c|c|c|c|c|}
    \hline Experiment & $l_{\text{min}},$ m   & $l_{\text{fid}},$ m & $\theta_{\text{min}},\theta_{\text{max}}$, mrad & $\Omega$, sr  \\ \hline
        \fas & 480  & 5 & $0,2.1$ & $1.3\cdot 10^{-5}$ \\ \hline
        \fac  & 101  & 18 & $1.6,4.1$ & $4.5\cdot 10^{-5}$ \\ \hline
    \end{tabular}
    \caption{Parameters of \fas and \fac configurations: the distance to the beginning of the decay volume $l_{\text{min}}$, the length of the decay volume $l_{\text{fid}}$, the polar angle coverage of detectors $\theta_{\text{min}},\theta_{\text{max}}$, and the solid angle $\Omega$ covered by the detectors.}
    \label{tab:parameters}
\end{table}

Let us compare the sensitivity of \fac and its modification at the lower bound (the regime $c\tau_{Y}\gamma_{Y}\gg l_{\text{max}}$) and the upper bound (regime $c\tau_{Y}\gamma_{Y}\lesssim l_{\text{min}}$) with the sensitivity of \fas. The lifetime scales as $\tau_{Y} \propto g_{Y}^{-2}$, where $g_{Y}$ is the mixing angle. The production branching ratio scales as $\text{Br}$

At the upper bound, the number of events behaves as $N_{\text{events}} \propto g^{2}_{Y}\times \exp[-l_{\text{min}}/c\tau_{Y}\gamma_{Y}]$. Therefore, neglecting the power $g^{2}_{Y}$, for ratio of the maximal probed mixing angles we get
\begin{equation}
    \frac{g^{2}_{\text{upper,\fac}}}{g^{2}_{\text{upper,\fas}}} \simeq \frac{l_{\text{min}}^{\text{\fas}}}{l_{\text{min}}^{\text{\fac}}} \cdot \frac{\langle \gamma_{Y}\rangle_{\text{\fac}}}{\langle \gamma_{Y}\rangle_{\text{\fas}}} \approx 5,
\end{equation}
given $\approx 5$ times smaller $l_{\text{min}}$ at \fac and $\langle \gamma_{Y}\rangle_{\text{\fac}}\approx \langle \gamma_{Y}\rangle_{\text{\fas}}$.

At the lower bound, $e^{-z/c\tau_{Y}\gamma_{Y}}\approx 1$, and the ratio of the probed mixing angles is 
\begin{equation}
    \frac{g^{2}_{\text{lower,\fac}}}{g^{2}_{\text{lower,\fas}}} \approx \left(\frac{\epsilon^{\text{\fas}}_{\text{geom}}}{\epsilon^{\text{\fac}}_{\text{geom}}}\times \frac{l_{\text{det}}^{\text{\fas}}}{l_{\text{det}}^{\text{\fac}}}\right)^{\kappa},
\end{equation}
Here, $\kappa = 1/2$ in the case when both $Y$ production and decay are controlled by $\theta$, and $\kappa = 1$ in the case when the production is controlled by different couplings. $\epsilon_{\text{geom}}$ is the averaged geometric acceptance at the lower bound:
\begin{equation}
    \epsilon_{\text{geom}}\approx \frac{1}{l_{\text{det}}}\int f_{\theta_{Y},E_{Y}} \cdot \epsilon_{\text{decay}}(\gamma_{Y},\theta_{Y},z) d\theta_{Y}dE_{Y}dz
    \label{eq:geom-acceptances}
\end{equation}

\subsection{Scalar portal}
\begin{figure}[!h]
    \centering
    \subfloat[]{\includegraphics[width=0.52\textwidth]{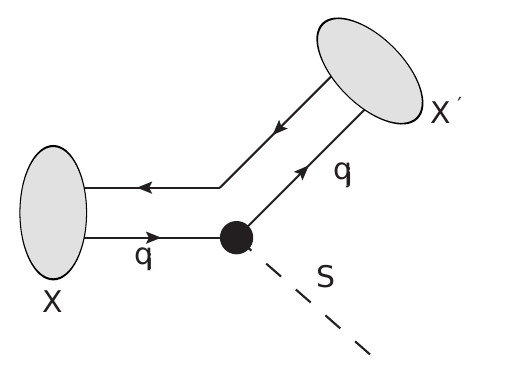}}~\subfloat[]{\includegraphics[width=0.46\textwidth]{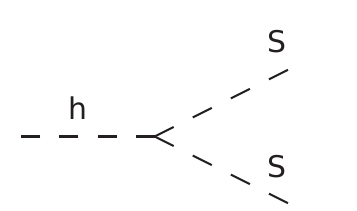}}   \\
    \subfloat[]{\raisebox{-0.55\height}{\includegraphics[width=0.52\textwidth]{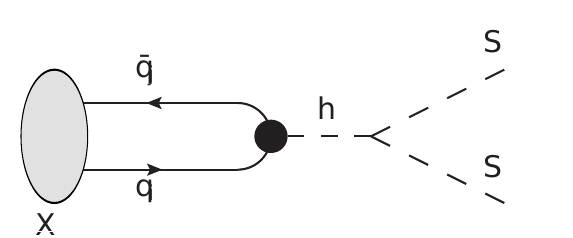}}~\raisebox{-0.5\height}{\includegraphics[width=0.46\textwidth]{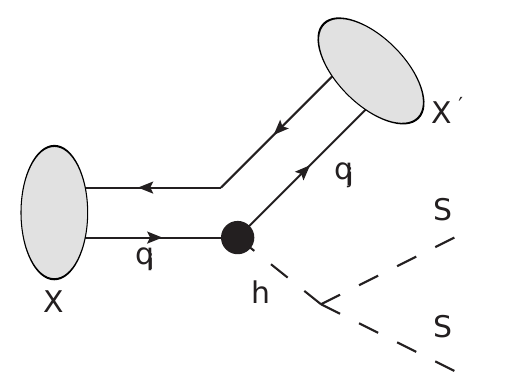}}}
    \caption{Diagrams of the production of the scalar $S$ in the model~\eqref{eq:scalar-portal}: meson decay $X\to X'+S$ (a) mediated by the mixing $\theta$; and the Higgs boson decay $h\to S+S$ (b), the mesons decays $X\to S+S$, $X\to X'+S+S$ (c) mediated by the quartic coupling $\alpha$.}
    \label{fig:scalar-production-diagrams}
\end{figure}

The production processes of scalars at the LHC are $h\to S+S$ for the Higgs bosons, and $B^{+/0}\to S+S+X_{s}$, $B_{s}\to S+S$, $B^{+/0}\to S+X_{s}$, for the $B$ mesons~\cite{Boiarska:2019jym}, see also Fig.~\ref{fig:scalar-production-diagrams}. The first three processes are mediated by the quartic coupling $\alpha$, while the process $B^{+/0}\to S+X_{s}$ by the mixing angle $\theta$.

\subsubsection{Geometric acceptance}
\label{sec:geom-acc}
Let us discuss the geometric acceptance. For the moment, we will drop the decay acceptance.

\begin{figure}
    \centering
    \includegraphics[width=0.6\textwidth]{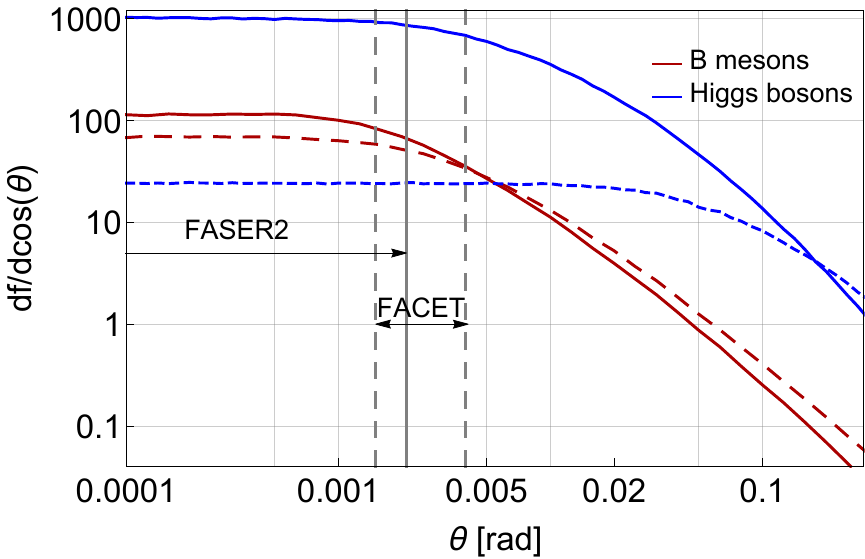}
    \caption{The solid angle distribution $df/d\cos(\theta)\sim df/d\Omega$ of B mesons, Higgs bosons (solid lines) and light scalars with $m_{S} = 50\text{ MeV}$ produced by their decays (dashed lines). The arrows indicate polar angle coverage of FASER2 and FACET experiments.}
    \label{fig:angular-distribution-mesons-higgs}
\end{figure}

The solid angle distribution $df/d\cos(\theta)\sim df/d\Omega$ of Higgs bosons, B mesons, and light scalars produced in their decays is shown in Fig.~\ref{fig:angular-distribution-mesons-higgs}.
The distribution of $B,h$ remains constant in the angular coverage of FASER2 and gradually drops by a factor 1.5-2 for the angular coverage of the FACET experiment. While the distribution of scalars with mass close to the kinematic threshold is the same as for their mother particles, the distribution of light scalars $m_{S}\ll m_{B},m_{h}/2$ gets broadened due to acquiring transverse momentum, of order of $p_{T}\simeq m_{B/h}/2$. Given that the typical $B,h$ energy in the far-forward direction is $\mathcal{O}(1\text{ TeV})$, the smearing is $\Delta \theta \sim p_{T}/1\text{ TeV}$ -- smaller than the angular coverage of \fac and \fas for scalars from $B$ mesons, and much larger for scalars from $h$. As a result, the angular distribution of light scalars from $B$ remains very similar to the distribution of $B$, while in the case of scalars from $h$ it is isotropic up to the angles $30\text{ mrad}$. This in particular suggests that \fac already has an optimal placement and size to search for particles from $B$ mesons.

Therefore, if not including the decay acceptance in Eq.~\eqref{eq:geom-acceptances}, for the geometric acceptance of scalars from $X = h,B$ one has
\begin{equation}
    \frac{\epsilon_{\text{geom},S}^{\fac}}{\epsilon_{\text{geom},S}^{\fas}} \approx \begin{cases} \Omega_{\fac}/\Omega_{\fas}, \quad m_{S}\ll m_{B},m_{h}/2\\ \epsilon_{\text{geom},X}^{\fac}/\epsilon_{\text{geom},X}^{\fas}, \quad m_{S}\to m_{B} \text{ or } m_{h}/2
    \end{cases}
\end{equation}

\begin{table}[!h]
    \centering
    \begin{tabular}{|c|c|c|c|c|}
    \hline \multirow{2}{*}{Experiment} & \multicolumn{2}{c|}{$\epsilon_{\text{geom},S},\ h\to SS$} & \multicolumn{2}{|c|}{$\epsilon_{\text{geom},S}, \ B\to KS$} \\ \cline{2-5}
        & $m_{S} = 50\text{ MeV}$ & $m_{S} = 62\text{ GeV}$ & $m_{S} = 50\text{ MeV}$ & $m_{S} = 5.1\text{ GeV}$  \\ \hline
    \fas & $5\cdot 10^{-5}$ & $1\cdot 10^{-3}$ & $2.8\cdot 10^{-3}$ & $7.5\cdot 10^{-3}$  \\ \hline
    \fac & $1.7\cdot 10^{-4}$ & $6.1\cdot 10^{-4}$ & $6.9\cdot 10^{-3}$ & $1\cdot 10^{-2}$  \\ \hline
    \end{tabular}
    \caption{Geometric acceptances (Eq.~\eqref{eq:geom-acceptances} for scalars produced by decays of Higgs bosons and $B$ mesons, for various choices of the scalar mass.}
    \label{tab:geom-acceptances-scalar}
\end{table}

Let us now discuss the effect of the decay acceptance. It becomes important if the characteristic angle between the decay products in a 2-body decay, $\langle\alpha\rangle \simeq 1\arcsin(2/\gamma_{S})$, exceeds the angle covered by the detector as seen from the beginning of the decay volume, which is $0.4\text{ rad}$ for \fas and $0.1\text{ rad}$ for \fac.

\begin{figure}
    \centering
    \includegraphics[width=0.5\textwidth]{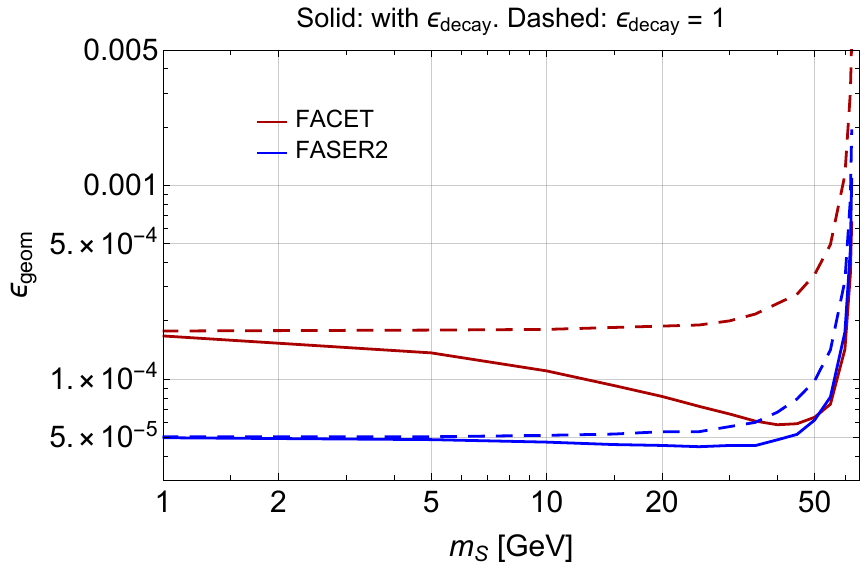}~\includegraphics[width=0.5\textwidth]{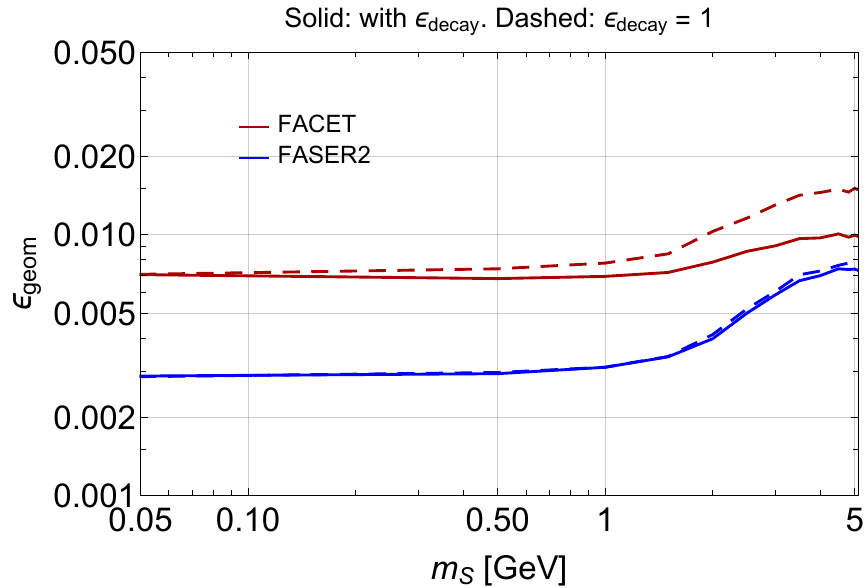}
    \caption{The behavior of the geometric acceptance~\eqref{eq:geom-acceptances} for scalars produced in decays $h\to SS$ and $B\to X_{s}+S$. The solid lines are obtained with taking the decay acceptance, $\epsilon_{\text{decay}}$, into account, whereas the dashed lines correspond to $\epsilon_{\text{decay}} = 1$.}
    \label{fig:geom-acceptances}
\end{figure}

Given the typical scalar energies $E_{S}= \mathcal{O}(1\text{ TeV})$, for light scalars with $m_{S}\ll m_{h/2},m_{B}$ the decay acceptance is 1. However, with the increase of the scalar mass, more and more decay products produced in the beginning of the decay volume fly in directions outside the detector coverage. This feature effectively shrinks $l_{\text{det}}$. For \fac, this effect is much more important than for \fas. As a result, in dependence on the scalar mass, the geometric acceptance at \fac drops even below the geometric acceptance at \fas for $h$ and becomes very close to the geometric acceptance at \fas for $B$, see Fig.~\ref{fig:geom-acceptances} and Table~\ref{tab:geom-acceptances-scalar}.

Given the ratio $l_{\text{det}}^{\text{\fas}}/l_{\text{det}}^{\text{\fac}} \approx 4$ and the behavior of the geometric acceptances (see Fig.~\ref{fig:geom-acceptances}), we conclude that the overall increase of the number of events in the regime of the lower bound at \fac as compared to \fas reaches a factor $2-15$ and $5-15$ for the case of the production from $h$ and $B$ respectively, in dependence on the scalar mass.

\subsubsection{Maximal number of events}
\label{sec:max-events}
In the case of the production from Higgs bosons, it is also useful to compare \textbf{the maximal possible number of events} at \fac and \fas. An estimate with an accuracy in a factor of two is given by
\begin{equation}
    N_{\text{events,max}} \approx N_{\text{h}}\cdot (2\cdot \text{Br}(h\to SS))\cdot \epsilon_{\text{geom}}\cdot P_{\text{decay,max}}, 
\end{equation}
where $\epsilon_{\text{geom}}$ is given by Eq.~\eqref{eq:geom-acceptances}, while $P_{\text{decay,max}}$ is the maximal value of the decay probability as a function of $l_{\text{decay,S}} = c\tau_{S}\gamma_{S}$, which depends only on $l_{\text{min}},l_{\text{max}}$:\footnote{In reality, the true maximal events number is even somewhat smaller, since for each given $m_{S},\theta^{2}$ there is the distribution in $l_{\text{decay}}$ due to the energy distribution of scalars.}
\begin{figure}
    \centering
    \includegraphics[width=0.6\textwidth]{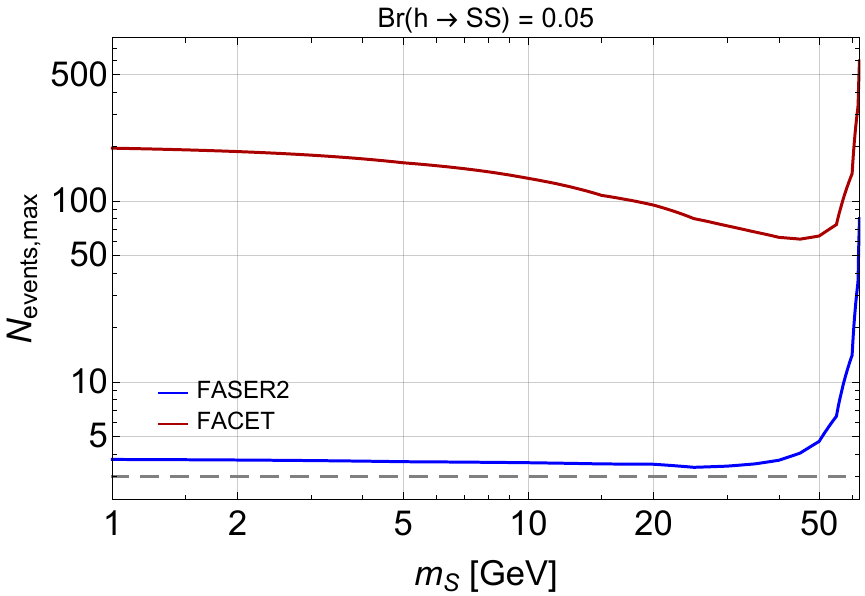}
    \caption{The maximal possible number of events (eq.~\eqref{eq:Nevents-max}) at \fas and \fac as a function of the scalar mass. The dashed gray line denotes 3 events defining the sensitivity domain of the experiments.}
    \label{fig:Nevents-max}
\end{figure}

\begin{equation}
   P_{\text{decay,max}}= \left[\left(\frac{l_{\text{min}}+l_{\text{fid}}}{l_{\text{min}}}\right)^{-\frac{l_{\text{min}}}{l_{\text{fid}}}} - \left(\frac{l_{\text{min}}+l_{\text{fid}}}{l_{\text{min}}}\right)^{-\frac{l_{\text{min}}+l_{\text{fid}}}{l_{\text{fid}}}} \right]\approx \begin{cases} 3.8\cdot 10^{-3}, \quad \text{\fas}, \\ 0.06, \quad \text{\fac} \end{cases}
\end{equation}
Plugging in all numbers, we get
\begin{equation}
    N_{\text{events,max}} \approx \begin{cases} 3.8\cdot \left(\frac{\epsilon_{\text{geom}}}{4.8\cdot 10^{-5}}\right), \quad \text{\fas}, \\ 200\cdot \left(\frac{\epsilon_{\text{geom}}}{1.8\cdot 10^{-4}}\right), \quad \text{\fac},  \end{cases}
    \label{eq:Nevents-max}
\end{equation}
see also Fig.~\ref{fig:Nevents-max}. 

From Eq.~\eqref{eq:Nevents-max}, we see that $N_{\text{events,max}}$ at \fas is very close to the number of events required at 95\% C.L. to observe one event in background free regime. More accurate estimates~\cite{Boiarska:2019vid} that included the energy distribution of scalars (which decreases the value of $P_{\text{decay,max}}$) showed that it is even lower, dropping below 3. This explains why \fas has no sensitivity to scalars from Higgs bosons in the domain $m_{S}\lesssim 45\text{ GeV}$.

\subsection{Comparison for HNLs}
\label{sec:HNLs}
Consider now the case of HNLs. The interaction vertices of HNLs with SM particles are similar to the vertices of active neutrinos $\nu_{\alpha}$, but are suppressed by the mixing angle $U_{\alpha}\ll 1$~\cite{Asaka:2005an,Asaka:2005pn}. 

At the LHC, the HNLs may be copiously produced in decays of $D,B$ mesons and $W$ bosons~\cite{Bondarenko:2018ptm}. In this Section, we consider HNLs that mix predominantly with $\nu_{e}$, keeping in mind that the results for the other mixings are similar.

\begin{figure}[h!]
    \centering
    \includegraphics[width=0.5\textwidth]{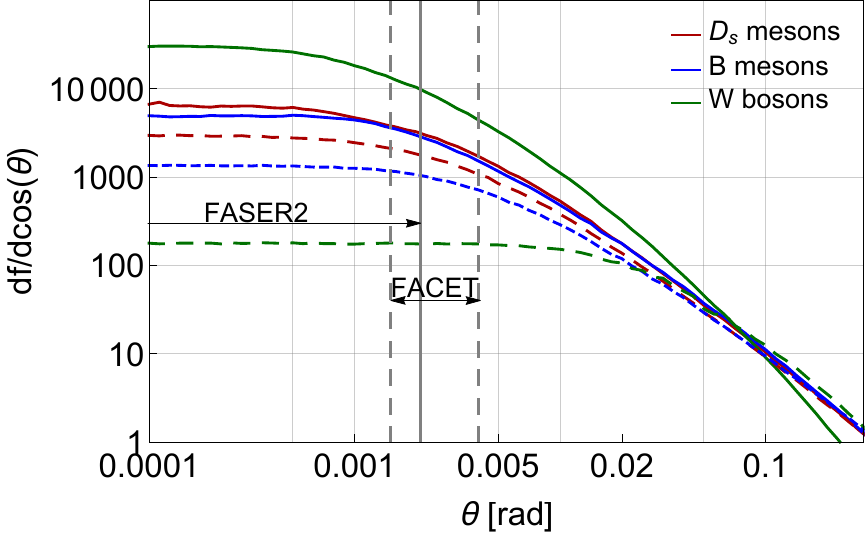}
    \caption{The angular distribution of $D_{s},B_{c}$ mesons, W bosons (solid lines), and light HNLs with $m_{N} = 50\text{ MeV}$ produced by their decays (dashed lines). The arrows indicate the polar angle coverage of FASER2 and FACET experiments.}
    \label{fig:angular-distribution-HNLs}
\end{figure}

For the qualitative comparison, we will consider the following production channels: $D_{s}\to N+e$, $B_{c}\to N+e$, $W\to N+e$, which respectively dominate the production of HNLs from $D$ mesons above $m_{N}\simeq 0.5\text{ GeV}$, from $B$ mesons above $m_{N}\simeq 3\text{ GeV}$, and from $W$ bosons. The angular distributions for these particles, as well as for light HNLs with mass $m_{N} = 50\text{ MeV}$ produced by their decays, are shown in Fig.~\ref{fig:angular-distribution-HNLs}.

The values of the geometric acceptances are given in Table~\ref{tab:geom-acceptances-HNL}.

\begin{table}[]
    \centering
    \begin{tabular}{|c|c|c|c|c|}
    \hline Experiment & $\epsilon_{\text{geom}}, D_{s}\to e+N$ & $\epsilon_{\text{geom}}, B_{c}\to e+N$  & $\epsilon_{\text{geom}}, W\to e+N $  \\ \hline
     \fas  & $8.5\cdot 10^{-3}$ & $4.9\cdot 10^{-3}$ & $4.2\cdot 10^{-4}$ \\ \hline
     \fac  & $1.4\cdot 10^{-2}$  & $1.3\cdot 10^{-2}$ & $1.3\cdot 10^{-3}$ \\ \hline
    \end{tabular}
    \caption{Geometric acceptances~\eqref{eq:geom-acceptances} for HNLs produced in decays of $D_{s},B_{c}$, and $W$ bosons (the decay acceptance is included). The HNL masses are $m_{N} = 1.5\text{ GeV}$ for the production from $D_{s}$, $3\text{ GeV}$ for the production from $B_{c}$, and $5\text{ GeV}$ for the production from $W$.}
    \label{tab:geom-acceptances-HNL}
\end{table}

The production from $W$ bosons is not important for the lower bound of the sensitivity of \fac and \fas. To demonstrate this, let us consider two mass ranges $m_{N}\lesssim m_{B_{c}}$ and $m_{N}>m_{B_{c}}$. In the mass range $m_{N} \lesssim m_{B_{c}}$, the production from $W$ competes with the production from $D$ and $B$. Let us compare the total number of HNLs produced by $B$ mesons and by $W$ bosons for \fac (for \fas, the situation is similar):
\begin{multline}
    \frac{N_{\text{prod}}^{\text{from W}}}{N_{\text{prod}}^{\text{from B}}} = \frac{N_{\text{W}}\cdot \epsilon_{\text{geom}}^{\text{from W}}}{N_{\text{B}}\cdot \epsilon_{\text{geom}}^{\text{from B}}}\times \frac{\text{Br}(W\to N+e)}{\sum_{i}f_{b\to B_{i}}\text{Br}(B_{i}\to N+X)} \simeq \\ \simeq 10^{-5}\times \frac{\text{Br}(W\to N+e)}{\sum_{i}f_{b\to B_{i}}\text{Br}(B_{i}\to N+X)} \ll 1
\end{multline}
where $f_{b\to B_{i}}$ is the fragmentation fraction of $b$ quark into a meson $B_{i}$, and we have taken into account that the second multiplier remains $\ll 10^{5}$ for practically all HNL masses below $m_{B_{c}}$~\cite{Bondarenko:2019yob}. 

For the mass range $m_{N}>m_{B_{c}}$, HNLs from $W$ are too short-lived and cannot reach the detector. Indeed, let us estimate the number of events with HNLs with mass $m_{B_{c}}$ produced by $W$ decays at $l_\text{decay}=c\tau_{N}\gamma_{N}\simeq l_{\text{min}}$. The number of events increases with decreasing $l_\text{decay}$ for $l_\text{decay} \gtrsim l_\text{min}$, so this should give an upper bound of events from $W$ for the lower bound of sensitivity. The corresponding mixing angle is $U^{2} \simeq l_{\text{min}}/c\tau_{N,U^{2}=1}\gamma_{N} \approx 6\cdot 10^{-8} $, where we used the results of~\cite{Bondarenko:2018ptm} for $\tau_{N}$ and $E_{N} =1\text{ TeV}$. The number of events is thus
\begin{equation}
    N_{N}^{(W)}\bigg|_{m_{N} = m_{B_{c}}}^{c\tau_{N}\gamma_{N} = l_{\text{min}}} =N_{W}\times \epsilon_{\text{geom},N}^{(W)}\times \text{Br}(W\to N)\times P_{\text{decay}}(c\tau_{N}\gamma_{N} = l_{\text{min}}) <1, 
\end{equation}
where we have used $\epsilon_{\text{geom},N}^{(W)}$ from Table~\ref{tab:geom-acceptances-HNL}.

\section{Results and discussion}
\label{sec:sensitivity-summary}
Using Eq.~\eqref{eq:nevents} and requiring $N_{\text{events}}>3$, corresponding to 95\% C.L. in the background free-regime of observing 1 event, we obtain the sensitivity of \fac and \fas to HNLs and Higgs-like scalars shown in Fig.~\ref{fig:sensitivity}.

The results agree with the estimates from Sec.~\ref{sec:analytic-estimates}. Namely, as compared to \fas, detectors of \fac covers $\simeq 3$ larger solid angle, while the decay volume of \fac is $\simeq 4$ times longer and located $\simeq 5$ times closer. Because of this, for dark scalars, \fac may probe the whole mass range $m_{S}<m_{h}/2$, while at \fas it is impossible to search for scalars in the mass range $m_{B}-m_{\pi} < m_{S}\lesssim 45\text{ GeV}$ due to the suppression of the geometric acceptance (see the discussion in Sec.~\ref{sec:max-events}). For HNLs, \fac may probe masses up to $m_{N}\simeq 6\text{ GeV}$, while \fas only up to $\simeq 4\text{ GeV}$, which is again due both to better sensitivity of \fac at the lower and upper bounds.

\begin{figure}[!h]
    \centering
    \includegraphics[width=0.5\textwidth]{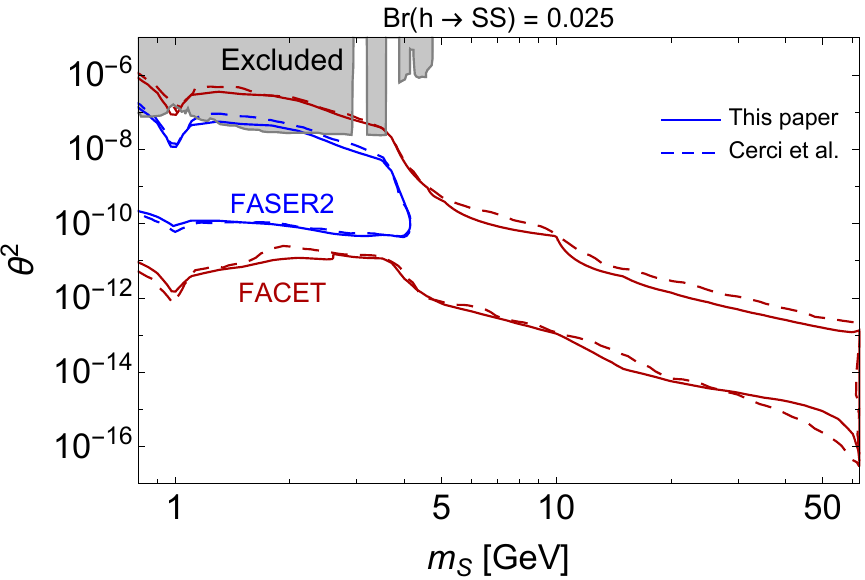}
    \caption{Comparison of the sensitivity of \fac to dark scalars obtained in our work and in~\cite{Cerci:2021nlb}, assuming $\text{Br}(h\to SS) = 0.025$ (for FACET) and $\text{Br}(h\to SS) = 0$ (for FASER2).}
    \label{fig:comparison-scalar}
\end{figure}

As a cross-check of our results, we compare the sensitivity to dark scalars obtained in this work with~\cite{Cerci:2021nlb}, which used FORESEE package~\cite{Kling:2021fwx}. Namely, we compared the sensitivities of FASER2 to scalars with zero quartic coupling, and the sensitivities of FACET assuming $\text{Br}(h\to SS) = 0.05$, see Fig.~\ref{fig:comparison-scalar}. The sensitivities agree well for low masses $m_{S}\lesssim 10\text{ GeV}$, but disagree by a factor of 2-3 at higher masses. The differences may be due to smaller decay width in~\cite{Cerci:2021nlb} (which explains the discrepancy at the upper bound) and the absence of the decay acceptance in their estimates.

An important feature shown in Fig.~\ref{fig:sensitivity} is that the sensitivity of \fac at the upper bound is better than the sensitivity of other experiments. The reason is the following: the upper bound is controlled by the ratio $\langle p\rangle/l_{\text{min}}$, where $\langle p\rangle$ is the mean momentum of decaying particles, and $l_{\text{min}}$ is the distance from the production point to the decay volume. While \fac has $l_{\text{min}}$ comparable to experiments such as SHiP and MATHUSLA, $\langle p\rangle$ is much higher -- $\simeq 10$ times higher than at SHiP, and $\simeq 100$ times higher than at MATHUSLA.

\section{Conclusions}
\label{sec:conclusions}
In this paper, we have estimated the potential of \fac, an experiment located in the far-forward direction at the LHC, to probe new physics, considering the models of scalar and fermion portals as an example. Using semi-analytic estimates, we have compared it with another proposed far-forward experiment, \fas, see Sec.~\ref{sec:analytic-estimates}. \fac has a larger decay volume, allowing it to probe the parameter space of long-lived particles, and is located closer to the interaction point (Table~\ref{tab:parameters}), which allows searching for short-lived particles. The combination of these features improves the sensitivity compared to \fas, significantly extending the probed mass range for both models, see Fig.~\ref{fig:sensitivity}. In particular, for dark scalars that are produced by decays of Higgs bosons, \fac may probe the whole kinematically allowed mass range $m_{S}<m_{h}/2$, whereas \fas has no sensitivity at scalar masses $m_{S}\lesssim 45\text{ GeV}$. For HNLs, \fac may probe masses up to the kinematic threshold for the production from $B_{c}$ mesons, while \fas has the sensitivity limited by $m_{N}\simeq 4\text{ GeV}$.

In addition, \fac is complementary to other LHC-based experiments (such as MATHUSLA) and SHiP. Indeed, the latter may search for new physics particles with much smaller mixing angles due to larger geometrical acceptance and decay volume. \fac, on the other hand, is better suited for probing particles with large couplings due to its on-axis placement: particles produced in the far-forward direction at the LHC have large $\gamma$ factors, which significantly increases their lifetime and makes it possible to reach the decay volume before decaying.

\section*{Acknowledgements} 
This project has received support from the European Union’s Horizon 2020 research and innovation programme under the Marie Sklodowska-Curie grant agreements No 860881-HIDDeN and 694896. KB is partly funded by the INFN PD51 INDARK grant.

\newpage 
\appendix 

\section{Mixing and quartic coupling at the lower bound of the sensitivity}

Below the $B$ meson mass, scalars may be produced by decays $h\to S+S, B\to X_{s}+S+S$, mediated by the quartic coupling, and $B\to X_{s}+S$ mediated by the mixing. Let us establish which of the production channels determines the lower bound of the sensitivity. Namely, let us compare the numbers of scalars produced by decays of $h$ and $B$ in the direction of the FACET experiment:
\begin{multline}
\frac{N_{\text{prod}}(B\to S)}{N_{\text{prod}}(h\to SS)} \sim \frac{N_{B}\cdot \chi^{(B)}_{S}\text{Br}(B\to S)\cdot \epsilon_{\text{geom},S}^{(B)}}{N_{h}\cdot \chi^{(h)}_{S}\text{Br}(h\to SS)\cdot \epsilon_{\text{geom},S}^{(h)}} \simeq \\  \simeq \begin{cases} \mathcal{O}(1)\cdot \frac{0.05}{\text{Br}(h\to SS)}\frac{\text{Br}(B\to X_{s}S)}{3\theta^{2}}\frac{\theta^{2}}{10^{-10}}, \quad B\to X_{s}S\\\mathcal{O}(10)\frac{\text{Br}(B\to X_{s}SS)}{5\cdot 10^{-10}}\quad B\to X_{s}SS, \end{cases}
\label{eq:processes-contributions}
\end{multline}
where we have normalized the branching ratios $\text{Br}(B\to X_{s}S), \text{Br}(B\to X_{s}SS)$ by their characteristic values, see~\cite{Boiarska:2019jym}. We conclude that below the kinematic threshold for $B\to X_{s}SS$, the lower bound is determined by the quartic production from $B$, whereas above there are two competing contributions from the mixing production from $B$ and quartic from $h$. 

\bibliographystyle{JHEP}
\bibliography{bib.bib}

\providecommand{\href}[2]{#2}\begingroup\raggedright\begin{thebibliography}{10}

\bibitem{Alekhin:2015byh}
S.~Alekhin et~al., {\it {A facility to Search for Hidden Particles at the CERN
  SPS: the SHiP physics case}},  {\em Rept. Prog. Phys.} {\bf 79} (2016),
  no.~12 124201, [\href{http://arxiv.org/abs/1504.04855}{{\tt
  arXiv:1504.04855}}].

\bibitem{Agrawal:2021dbo}
P.~Agrawal et~al., {\it {Feebly-interacting particles: FIPs 2020 workshop
  report}},  {\em Eur. Phys. J. C} {\bf 81} (2021), no.~11 1015,
  [\href{http://arxiv.org/abs/2102.12143}{{\tt arXiv:2102.12143}}].

\bibitem{SHiP:2015vad}
{\bf SHiP} Collaboration, M.~Anelli et~al., {\it {A facility to Search for
  Hidden Particles (SHiP) at the CERN SPS}},
  \href{http://arxiv.org/abs/1504.04956}{{\tt arXiv:1504.04956}}.

\bibitem{DUNE:2015lol}
{\bf DUNE} Collaboration, R.~Acciarri et~al., {\it {Long-Baseline Neutrino
  Facility (LBNF) and Deep Underground Neutrino Experiment (DUNE)}: {Conceptual
  Design Report, Volume 2: The Physics Program for DUNE at LBNF}},
  \href{http://arxiv.org/abs/1512.06148}{{\tt arXiv:1512.06148}}.

\bibitem{Baldini:2021hfw}
W.~Baldini et~al., {\it {SHADOWS (Search for Hidden And Dark Objects With the
  SPS)}},  \href{http://arxiv.org/abs/2110.08025}{{\tt arXiv:2110.08025}}.

\bibitem{NA62:2017rwk}
{\bf NA62} Collaboration, E.~Cortina~Gil et~al., {\it {The Beam and detector of
  the NA62 experiment at CERN}},  {\em JINST} {\bf 12} (2017), no.~05 P05025,
  [\href{http://arxiv.org/abs/1703.08501}{{\tt arXiv:1703.08501}}].

\bibitem{Curtin:2018mvb}
D.~Curtin et~al., {\it {Long-Lived Particles at the Energy Frontier: The
  MATHUSLA Physics Case}},  {\em Rept. Prog. Phys.} {\bf 82} (2019), no.~11
  116201, [\href{http://arxiv.org/abs/1806.07396}{{\tt arXiv:1806.07396}}].

\bibitem{Aielli:2019ivi}
G.~Aielli et~al., {\it {Expression of interest for the CODEX-b detector}},
  {\em Eur. Phys. J. C} {\bf 80} (2020), no.~12 1177,
  [\href{http://arxiv.org/abs/1911.00481}{{\tt arXiv:1911.00481}}].

\bibitem{Bauer:2019vqk}
M.~Bauer, O.~Brandt, L.~Lee, and C.~Ohm, {\it {ANUBIS: Proposal to search for
  long-lived neutral particles in CERN service shafts}},
  \href{http://arxiv.org/abs/1909.13022}{{\tt arXiv:1909.13022}}.

\bibitem{Gligorov:2018vkc}
V.~V. Gligorov, S.~Knapen, B.~Nachman, M.~Papucci, and D.~J. Robinson, {\it
  {Leveraging the ALICE/L3 cavern for long-lived particle searches}},  {\em
  Phys. Rev. D} {\bf 99} (2019), no.~1 015023,
  [\href{http://arxiv.org/abs/1810.03636}{{\tt arXiv:1810.03636}}].

\bibitem{FASER:2018bac}
{\bf FASER} Collaboration, A.~Ariga et~al., {\it {Technical Proposal for FASER:
  ForwArd Search ExpeRiment at the LHC}},
  \href{http://arxiv.org/abs/1812.09139}{{\tt arXiv:1812.09139}}.

\bibitem{FASER:2019aik}
{\bf FASER} Collaboration, A.~Ariga et~al., {\it {FASER: ForwArd Search
  ExpeRiment at the LHC}},  \href{http://arxiv.org/abs/1901.04468}{{\tt
  arXiv:1901.04468}}.

\bibitem{FASER:2019dxq}
{\bf FASER} Collaboration, H.~Abreu et~al., {\it {Detecting and Studying
  High-Energy Collider Neutrinos with FASER at the LHC}},  {\em Eur. Phys. J.
  C} {\bf 80} (2020), no.~1 61, [\href{http://arxiv.org/abs/1908.02310}{{\tt
  arXiv:1908.02310}}].

\bibitem{FASER:2020gpr}
{\bf FASER} Collaboration, H.~Abreu et~al., {\it {Technical Proposal:
  FASERnu}},  \href{http://arxiv.org/abs/2001.03073}{{\tt arXiv:2001.03073}}.

\bibitem{SHiP:2020sos}
{\bf SHiP} Collaboration, C.~Ahdida et~al., {\it {SND@LHC}},
  \href{http://arxiv.org/abs/2002.08722}{{\tt arXiv:2002.08722}}.

\bibitem{Feng:2022inv}
J.~L. Feng et~al., {\it {The Forward Physics Facility at the High-Luminosity
  LHC}},  \href{http://arxiv.org/abs/2203.05090}{{\tt arXiv:2203.05090}}.

\bibitem{Cerci:2021nlb}
S.~Cerci et~al., {\it {FACET: A new long-lived particle detector in the very
  forward region of the CMS experiment}},
  \href{http://arxiv.org/abs/2201.00019}{{\tt arXiv:2201.00019}}.

\bibitem{Du:2021cmt}
M.~Du, R.~Fang, Z.~Liu, and V.~Q. Tran, {\it {Enhanced long-lived dark photon
  signals at lifetime frontier detectors}},  {\em Phys. Rev. D} {\bf 105}
  (2022), no.~5 055012, [\href{http://arxiv.org/abs/2111.15503}{{\tt
  arXiv:2111.15503}}].

\bibitem{Liu:2022ugx}
W.~Liu, J.~Li, J.~Li, and H.~Sun, {\it {Testing the seesaw mechanisms via
  displaced right-handed neutrinos from a light scalar at the HL-LHC}},  {\em
  Phys. Rev. D} {\bf 106} (2022), no.~1 015019,
  [\href{http://arxiv.org/abs/2204.03819}{{\tt arXiv:2204.03819}}].

\bibitem{Kachanovich:2020yhi}
A.~Kachanovich, U.~Nierste, and I.~Ni\v{s}and\v{z}i\'c, {\it {Higgs portal to
  dark matter and $B\to K^{(*)}$ decays}},  {\em Eur. Phys. J. C} {\bf 80}
  (2020), no.~7 669, [\href{http://arxiv.org/abs/2003.01788}{{\tt
  arXiv:2003.01788}}].

\bibitem{Filimonova:2019tuy}
A.~Filimonova, R.~Sch\"afer, and S.~Westhoff, {\it {Probing dark sectors with
  long-lived particles at BELLE II}},  {\em Phys. Rev. D} {\bf 101} (2020),
  no.~9 095006, [\href{http://arxiv.org/abs/1911.03490}{{\tt
  arXiv:1911.03490}}].

\bibitem{Drewes:2019fou}
M.~Drewes and J.~Hajer, {\it {Heavy Neutrinos in displaced vertex searches at
  the LHC and HL-LHC}},  {\em JHEP} {\bf 02} (2020) 070,
  [\href{http://arxiv.org/abs/1903.06100}{{\tt arXiv:1903.06100}}].

\bibitem{Abdullahi:2022jlv}
A.~M. Abdullahi et~al., {\it {The Present and Future Status of Heavy Neutral
  Leptons}},  in {\em {2022 Snowmass Summer Study}}, 3, 2022.
\newblock \href{http://arxiv.org/abs/2203.08039}{{\tt arXiv:2203.08039}}.

\bibitem{Bezrukov:2009yw}
F.~Bezrukov and D.~Gorbunov, {\it {Light inflaton Hunter's Guide}},  {\em JHEP}
  {\bf 05} (2010) 010, [\href{http://arxiv.org/abs/0912.0390}{{\tt
  arXiv:0912.0390}}].

\bibitem{Boiarska:2019jym}
I.~Boiarska, K.~Bondarenko, A.~Boyarsky, V.~Gorkavenko, M.~Ovchynnikov, and
  A.~Sokolenko, {\it {Phenomenology of GeV-scale scalar portal}},  {\em JHEP}
  {\bf 11} (2019) 162, [\href{http://arxiv.org/abs/1904.10447}{{\tt
  arXiv:1904.10447}}].

\bibitem{Bird:2004ts}
C.~Bird, P.~Jackson, R.~V. Kowalewski, and M.~Pospelov, {\it {Search for dark
  matter in $b\to s$ transitions with missing energy}},  {\em Phys. Rev. Lett.}
  {\bf 93} (2004) 201803, [\href{http://arxiv.org/abs/hep-ph/0401195}{{\tt
  hep-ph/0401195}}].

\bibitem{Batell:2009jf}
B.~Batell, M.~Pospelov, and A.~Ritz, {\it {Multi-lepton Signatures of a Hidden
  Sector in Rare B Decays}},  {\em Phys. Rev.} {\bf D83} (2011) 054005,
  [\href{http://arxiv.org/abs/0911.4938}{{\tt arXiv:0911.4938}}].

\bibitem{Clarke:2013aya}
J.~D. Clarke, R.~Foot, and R.~R. Volkas, {\it {Phenomenology of a very light
  scalar (100 MeV $\le m_h\le$ 10 GeV) mixing with the SM Higgs}},  {\em JHEP}
  {\bf 02} (2014) 123, [\href{http://arxiv.org/abs/1310.8042}{{\tt
  arXiv:1310.8042}}].

\bibitem{Schmidt-Hoberg:2013hba}
K.~Schmidt-Hoberg, F.~Staub, and M.~W. Winkler, {\it {Constraints on light
  mediators: confronting dark matter searches with B physics}},  {\em Phys.
  Lett.} {\bf B727} (2013) 506--510,
  [\href{http://arxiv.org/abs/1310.6752}{{\tt arXiv:1310.6752}}].

\bibitem{Evans:2017lvd}
J.~A. Evans, {\it {Detecting Hidden Particles with MATHUSLA}},  {\em Phys.
  Rev.} {\bf D97} (2018), no.~5 055046,
  [\href{http://arxiv.org/abs/1708.08503}{{\tt arXiv:1708.08503}}].

\bibitem{Bezrukov:2018yvd}
F.~Bezrukov, D.~Gorbunov, and I.~Timiryasov, {\it {Uncertainties of hadronic
  scalar decay calculations}},  \href{http://arxiv.org/abs/1812.08088}{{\tt
  arXiv:1812.08088}}.

\bibitem{Monin:2018lee}
A.~Monin, A.~Boyarsky, and O.~Ruchayskiy, {\it {Hadronic decays of a light
  Higgs-like scalar}},  {\em Phys. Rev.} {\bf D99} (2019), no.~1 015019,
  [\href{http://arxiv.org/abs/1806.07759}{{\tt arXiv:1806.07759}}].

\bibitem{Winkler:2018qyg}
M.~W. Winkler, {\it {Decay and detection of a light scalar boson mixing with
  the Higgs boson}},  {\em Phys. Rev.} {\bf D99} (2019), no.~1 015018,
  [\href{http://arxiv.org/abs/1809.01876}{{\tt arXiv:1809.01876}}].

\bibitem{Frugiuele:2018coc}
C.~Frugiuele, E.~Fuchs, G.~Perez, and M.~Schlaffer, {\it {Relaxion and light
  (pseudo)scalars at the HL-LHC and lepton colliders}},  {\em JHEP} {\bf 10}
  (2018) 151, [\href{http://arxiv.org/abs/1807.10842}{{\tt arXiv:1807.10842}}].

\bibitem{Helmboldt:2016zns}
A.~J. Helmboldt and M.~Lindner, {\it {Prospects for three-body Higgs boson
  decays into extra light scalars}},  {\em Phys. Rev.} {\bf D95} (2017), no.~5
  055008, [\href{http://arxiv.org/abs/1609.08127}{{\tt arXiv:1609.08127}}].

\bibitem{Voloshin:1985tc}
M.~B. Voloshin, {\it {Once Again About the Role of Gluonic Mechanism in
  Interaction of Light Higgs Boson with Hadrons}},  {\em Sov. J. Nucl. Phys.}
  {\bf 44} (1986) 478. [Yad. Fiz.44,738(1986)].

\bibitem{Raby:1988qf}
S.~Raby and G.~B. West, {\it {The Branching Ratio for a Light Higgs to Decay
  Into $\mu^+ \mu^-$ Pairs}},  {\em Phys. Rev.} {\bf D38} (1988) 3488.

\bibitem{Truong:1989my}
T.~N. Truong and R.~S. Willey, {\it {Branching Ratios for Decays of Light Higgs
  Bosons}},  {\em Phys. Rev.} {\bf D40} (1989) 3635.

\bibitem{Donoghue:1990xh}
J.~F. Donoghue, J.~Gasser, and H.~Leutwyler, {\it {The Decay of a Light Higgs
  Boson}},  {\em Nucl. Phys.} {\bf B343} (1990) 341--368.

\bibitem{Willey:1982ti}
R.~S. Willey and H.~L. Yu, {\it {The Decays $K^\pm \to \pi^\pm \ell^+ \ell^-$
  and Limits on the Mass of the Neutral Higgs Boson}},  {\em Phys. Rev.} {\bf
  D26} (1982) 3287.

\bibitem{Willey:1986mj}
R.~S. Willey, {\it {Limits on Light Higgs Bosons From the Decays $K^\pm \to
  \pi^\pm \ell^- \ell^+$}},  {\em Phys. Lett.} {\bf B173} (1986) 480--484.

\bibitem{Grzadkowski:1983yp}
B.~Grzadkowski and P.~Krawczyk, {\it {HIGGS PARTICLE EFFECTS IN FLAVOR CHANGING
  TRANSITIONS}},  {\em Z. Phys.} {\bf C18} (1983) 43--45.

\bibitem{Leutwyler:1989xj}
H.~Leutwyler and M.~A. Shifman, {\it {Light Higgs Particle in Decays of $K$ and
  $\eta$ Mesons}},  {\em Nucl. Phys.} {\bf B343} (1990) 369--397.

\bibitem{Haber:1987ua}
H.~E. Haber, A.~S. Schwarz, and A.~E. Snyder, {\it {Hunting the Higgs in $B$
  Decays}},  {\em Nucl. Phys.} {\bf B294} (1987) 301--320.

\bibitem{Chivukula:1988gp}
R.~S. Chivukula and A.~V. Manohar, {\it {LIMITS ON A LIGHT HIGGS BOSON}},  {\em
  Phys. Lett.} {\bf B207} (1988) 86. [Erratum: Phys. Lett.B217,568(1989)].

\bibitem{Sirunyan:2018owy}
{\bf CMS} Collaboration, A.~M. Sirunyan et~al., {\it {Search for invisible
  decays of a Higgs boson produced through vector boson fusion in proton-proton
  collisions at $\sqrt{s} =$ 13 TeV}},  {\em Phys. Lett.} {\bf B793} (2019)
  520--551, [\href{http://arxiv.org/abs/1809.05937}{{\tt arXiv:1809.05937}}].

\bibitem{Aaboud:2018sfi}
{\bf ATLAS} Collaboration, M.~Aaboud et~al., {\it {Search for invisible Higgs
  boson decays in vector boson fusion at $\sqrt{s} = 13$ TeV with the ATLAS
  detector}},  {\em Phys. Lett.} {\bf B793} (2019) 499--519,
  [\href{http://arxiv.org/abs/1809.06682}{{\tt arXiv:1809.06682}}].

\bibitem{Bechtle:2014ewa}
P.~Bechtle, S.~Heinemeyer, O.~Stål, T.~Stefaniak, and G.~Weiglein, {\it
  {Probing the Standard Model with Higgs signal rates from the Tevatron, the
  LHC and a future ILC}},  {\em JHEP} {\bf 11} (2014) 039,
  [\href{http://arxiv.org/abs/1403.1582}{{\tt arXiv:1403.1582}}].

\bibitem{Boiarska:2019vid}
I.~Boiarska, K.~Bondarenko, A.~Boyarsky, M.~Ovchynnikov, O.~Ruchayskiy, and
  A.~Sokolenko, {\it {Light scalar production from Higgs bosons and FASER 2}},
  {\em JHEP} {\bf 05} (2020) 049, [\href{http://arxiv.org/abs/1908.04635}{{\tt
  arXiv:1908.04635}}].

\bibitem{Bondarenko:2019yob}
K.~Bondarenko, A.~Boyarsky, M.~Ovchynnikov, and O.~Ruchayskiy, {\it
  {Sensitivity of the intensity frontier experiments for neutrino and scalar
  portals: analytic estimates}},  {\em JHEP} {\bf 08} (2019) 061,
  [\href{http://arxiv.org/abs/1902.06240}{{\tt arXiv:1902.06240}}].

\bibitem{Bondarenko:2018ptm}
K.~Bondarenko, A.~Boyarsky, D.~Gorbunov, and O.~Ruchayskiy, {\it {Phenomenology
  of GeV-scale Heavy Neutral Leptons}},  {\em JHEP} {\bf 11} (2018) 032,
  [\href{http://arxiv.org/abs/1805.08567}{{\tt arXiv:1805.08567}}].

\bibitem{Workman:2022ynf}
{\bf Particle Data Group} Collaboration, R.~L. Workman and Others, {\it {Review
  of Particle Physics}},  {\em PTEP} {\bf 2022} (2022) 083C01.

\bibitem{Cepeda:2019klc}
M.~Cepeda et~al., {\it {Report from Working Group 2}: {Higgs Physics at the
  HL-LHC and HE-LHC}},  {\em CERN Yellow Rep. Monogr.} {\bf 7} (2019) 221--584,
  [\href{http://arxiv.org/abs/1902.00134}{{\tt arXiv:1902.00134}}].

\bibitem{Cacciari:1998it}
M.~Cacciari, M.~Greco, and P.~Nason, {\it {The P(T) spectrum in heavy flavor
  hadroproduction}},  {\em JHEP} {\bf 05} (1998) 007,
  [\href{http://arxiv.org/abs/hep-ph/9803400}{{\tt hep-ph/9803400}}].

\bibitem{Cacciari:2001td}
M.~Cacciari, S.~Frixione, and P.~Nason, {\it {The p(T) spectrum in heavy flavor
  photoproduction}},  {\em JHEP} {\bf 03} (2001) 006,
  [\href{http://arxiv.org/abs/hep-ph/0102134}{{\tt hep-ph/0102134}}].

\bibitem{Cacciari:2012ny}
M.~Cacciari, S.~Frixione, N.~Houdeau, M.~L. Mangano, P.~Nason, and G.~Ridolfi,
  {\it {Theoretical predictions for charm and bottom production at the LHC}},
  {\em JHEP} {\bf 10} (2012) 137, [\href{http://arxiv.org/abs/1205.6344}{{\tt
  arXiv:1205.6344}}].

\bibitem{Cacciari:2015fta}
M.~Cacciari, M.~L. Mangano, and P.~Nason, {\it {Gluon PDF constraints from the
  ratio of forward heavy-quark production at the LHC at $\sqrt{S}=7$ and 13
  TeV}},  {\em Eur. Phys. J.} {\bf C75} (2015), no.~12 610,
  [\href{http://arxiv.org/abs/1507.06197}{{\tt arXiv:1507.06197}}].

\bibitem{ATLAS:2016nlr}
{\bf ATLAS} Collaboration, M.~Aaboud et~al., {\it {Measurement of the
  $W^{\pm}Z$ boson pair-production cross section in $pp$ collisions at
  $\sqrt{s}=13$ TeV with the ATLAS Detector}},  {\em Phys. Lett. B} {\bf 762}
  (2016) 1--22, [\href{http://arxiv.org/abs/1606.04017}{{\tt
  arXiv:1606.04017}}].

\bibitem{Kling:2021fwx}
F.~Kling and S.~Trojanowski, {\it {Forward experiment sensitivity estimator for
  the LHC and future hadron colliders}},  {\em Phys. Rev. D} {\bf 104} (2021),
  no.~3 035012, [\href{http://arxiv.org/abs/2105.07077}{{\tt
  arXiv:2105.07077}}].

\bibitem{Asaka:2005an}
T.~Asaka, S.~Blanchet, and M.~Shaposhnikov, {\it {The nuMSM, dark matter and
  neutrino masses}},  {\em Phys. Lett. B} {\bf 631} (2005) 151--156,
  [\href{http://arxiv.org/abs/hep-ph/0503065}{{\tt hep-ph/0503065}}].

\bibitem{Asaka:2005pn}
T.~Asaka and M.~Shaposhnikov, {\it {The $\nu$MSM, dark matter and baryon
  asymmetry of the universe}},  {\em Phys. Lett. B} {\bf 620} (2005) 17--26,
  [\href{http://arxiv.org/abs/hep-ph/0505013}{{\tt hep-ph/0505013}}].

\end{thebibliography}\endgroup
\end{document}